\documentclass{aa}
\usepackage[varg]{txfonts}

\usepackage{epsfig,graphicx,natbib,url,twoopt}
\usepackage{enumitem}
\usepackage[
  breaklinks = true,
  colorlinks = true,
  urlcolor   = blue,
  citecolor  = blue,
  linkcolor  = blue,
]{hyperref}
\usepackage{float}
\setcitestyle{citesep={,}} 

\def\Halpha{\mbox{H\hspace{0.1ex}$\alpha$}}
\def\Hbeta{\mbox{H\hspace{0.1ex}$\beta$}}
\def\Hepsilon{\mbox{H\hspace{0.1ex}$\varepsilon$}}

\def\FeII{\ion{Fe}{ii}}

\def\HeI{\ion{He}{i}}

\def\NiII{\ion{Ni}{ii}}

\def\CaIR{\ion{Ca}{ii}~8542\,\AA}
\def\CaK{\ion{Ca}{ii}~K}
\def\CaH{\ion{Ca}{ii}~H}

\def\Mgtriplet{\ion{Mg}{ii}~triplet}
\def\Siiv{\ion{Si}{iv}}

\defcitealias{2024A&A...689A.156B}{Paper I}
\defcitealias{2025A&A...693A.221B}{Paper II}

\begin{document} 
\title{Small-scale dynamic phenomena associated with interacting fan-spine topologies: quiet-Sun Ellerman bombs, UV brightenings, and chromospheric inverted-Y-shaped jets}

\titlerunning{Fan-spine topologies: QSEBs, UV brightenings, and chromospheric inverted-Y-shaped jets}

\author{Aditi Bhatnagar \inst{1,2}
\and Avijeet Prasad \inst{1,2}
\and Daniel Nóbrega-Siverio \inst{3,4,1,2}
\and Luc Rouppe van der Voort \inst{1,2}
\and Jayant Joshi \inst{5}}      

\institute{
  Institute of Theoretical Astrophysics,
  University of Oslo, %
  P.O. Box 1029 Blindern, N-0315 Oslo, Norway
\and
  Rosseland Centre for Solar Physics,
  University of Oslo, %
  P.O. Box 1029 Blindern, N-0315 Oslo, Norway
\and
  Instituto de Astrofísica de Canarias, 
  38205 La Laguna, Tenerife, Spain
\and
  Universidad de La Laguna, Dept. Astrofísica, 
  38206 La Laguna, Tenerife, Spain
\and
  Indian Institute of Astrophysics, 
  II Block, Koramangala, Bengaluru 560 034, India
}


\date{Submitted to A\&A on 7 December 2024 / Accepted 23 April 2025 }  

\abstract
{ 
Quiet-Sun Ellerman bombs (QSEBs) are small-scale magnetic reconnection events in the lower solar atmosphere. Sometimes, they exhibit transition region counterparts, known as ultraviolet (UV) brightenings. Magnetic field extrapolations suggest that QSEBs can occur at various locations of a fan-spine topology, with UV brightening occurring at the magnetic null point through a common reconnection process.}
{ 
We aim to understand how more complex magnetic field configurations such as interacting fan-spine topologies can cause small-scale dynamic phenomena in the lower atmosphere.}
{ 
QSEBs were detected using \textit{k}-means clustering on \Hbeta\ observations from the Swedish 1-m Solar Telescope (SST). Further, chromospheric inverted-Y-shaped jets were identified in the \Hbeta\ blue wing. Magnetic field topologies were determined through potential field extrapolations from photospheric magnetograms derived from spectro-polarimetric observations in the \ion{Fe}{i} 6173~\AA\ line. UV brightenings were detected in IRIS 1400~\AA\ slit-jaw images.}
{ 
We identify two distinct magnetic configurations associated with QSEBs, UV brightenings, and chromospheric inverted-Y-shaped jets. The first involves a nested fan-spine structure where, due to flux emergence, an inner 3D null forms inside the fan surface of an outer 3D null with some overlap. The QSEBs occur at two footpoints along the shared fan surface, with the UV brightening located near the outer 3D null point. The jet originates close to the two QSEBs and follows the path of high squashing factor $Q$. We discuss a comparable scenario using a 2D numerical experiment with the Bifrost code. In the second case, two adjacent fan-spine topologies share fan footpoints at a common positive polarity patch, with the QSEB, along with a chromospheric inverted-Y-shaped jet, occurring at the intersection having high $Q$ values. The width of the jets in our examples is about 0\farcs3, and the height varies between 1\arcsec --2\arcsec. The width of the cusp measures between 1\arcsec --2\arcsec.}
{ 
This study demonstrates through observational and modelling support that small-scale dynamic phenomena, such as associated QSEBs, UV brightenings, and chromospheric inverted-Y-shaped jets share a common origin driven by magnetic reconnection between interacting fan-spine topologies.}


\maketitle

\section{Introduction}
\label{sec:introduction}
Ellerman Bombs (EBs) are short-lived, small-scale brightenings observed in solar active regions. 
They were first observed in the wings of the \Halpha\ spectral line at 6563~\AA\ \citep{1917ApJ....46..298E}, and are characterised by their moustache-shaped spectral profile \citep{1964ARA&A...2..363S}, with enhanced emissions in the wings and an unaffected line core.
Ellerman bombs are driven by magnetic reconnection in the photosphere, often in connection with magnetic flux emergence. 
They exhibit a flame-like morphology when observed close to the solar limb and have lifetimes ranging from a few seconds to minutes \citep[e.g.,][]{1982SoPh...79...77K, 1998SoPh..182..381N, 2011ApJ...736...71W, 2013JPhCS.440a2007R, 2015ApJ...798...19N}. 
Similar events are also observed in the quieter regions of the Sun and are known as quiet-Sun Ellerman bombs 
\citep[QSEB,][]{2016A&A...592A.100R}.
QSEBs are ubiquitous in nature \citep{2020A&A...641L...5J, 2022A&A...664A..72J}, as inferred from studies based on high resolution \Hbeta\ observations from the Swedish 1-m Solar Telescope \citep[SST,][]{2003SPIE.4853..341S}. 
The current estimate is that around 750\,000 QSEBs are present at any given time on the sun, which was obtained using higher resolution \Hepsilon\ observations from SST \citep{2024A&A...683A.190R}.

Numerous topological scenarios have been proposed for EB formation. 
EBs can occur due to magnetic reconnection between the newly emerging flux and pre-existing magnetic fields \citep[e.g.,][]{2008ApJ...684..736W, 2010PASJ...62..879H, 2017ApJ...839...22H, 2024A&A...686A.218N}. 
They are also observed in unipolar regions \citep{2002ApJ...575..506G, 2008ApJ...684..736W, 2010PASJ...62..879H}, where a misalignment of magnetic field lines can lead to the formation of quasi-separatrix layers 
\citep[QSL,][]{1996A&A...308..643D}.  
EBs can also occur at bald patches, which are in regions with U-shaped photospheric magnetic loops, as proposed by \citet{2004ApJ...614.1099P, 2006AdSpR..38..902P, 2012ASPC..455..177P, 2012EAS....55..115P}.
%

Recent studies suggest a strong connection between QSEBs and chromospheric dynamics. \citet{2023ApJ...944..171B} demonstrated that flux emergence increased the chromospheric spicule activity while also driving reconnection in the lower atmosphere leading to QSEBs. 
%
\citet{2025arXiv250405396S} 
found a large number of QSEBs that could be connected to the formation of spicules.
Spicules are thin, jet-like excursions of chromospheric plasma that are ubiquitous in the chromosphere and are classified as Type I and Type II \citep{2007PASJ...59S.655D}.
Type I spicules are driven by magnetoacoustic shocks originating from photospheric oscillations and convection \citep{2006ApJ...647L..73H, 2007ApJ...655..624D}.
The formation of Type II spicules has been linked to the release of built-up magnetic tension, as demonstrated by radiative-MHD simulations \citep{2017ApJ...847...36M, 2017Sci...356.1269M, 2020ApJ...889...95M}.
They have been studied using on-disk observations in many different spectral lines like \Halpha, \Hbeta, \CaIR, and \CaK\ and have been termed as Rapid blue-shifted and red-shifted excursions \citep[RBEs and RREs,][]{2008ApJ...679L.167L, 2009ApJ...705..272R, 2012ApJ...752..108S, 2019A&A...631L...5B}, which are found close to strong network regions with enhanced magnetic fields. 
Type II spicules are believed to form via magnetic reconnection \citep[see, e.g.,][]{2019Sci...366..890S}.
The reconnection between emerging and pre-existing magnetic fields during flux emergence can also lead to other phenomena such as hot jets and cool surges \citep{1996PASJ...48..353Y, 2008ApJ...683L..83N, 2016ApJ...822...18N}. 
Many of these jets exhibit an inverted-Y-shaped structure. \citet{2007Sci...318.1591S} found similar structures in the \CaH\ line in SOT/Hinode \citep{2007SoPh..243....3K, 2008SoPh..249..167T} images and named them chromospheric anemone jets. 
\citet{2018A&A...611L...6P} observed inverted-Y-shaped jets in \Halpha\ above low-lying transition region loops, attributing their formation to magnetic reconnection.
Inverted-Y-shaped jets have also been observed by \citet{2011ApJ...736L..35Y} who reported them in intergranular lanes. 
Recently, the work of \cite{2024A&A...686A.218N} highlighted a sequence of events, where small-scale magnetic flux emergence in a relatively quiet region first triggered EBs, followed by associated ultraviolet (UV) bursts and surges.

\begin{figure*}[!ht]
    \centering
    \includegraphics[width=0.76\linewidth]{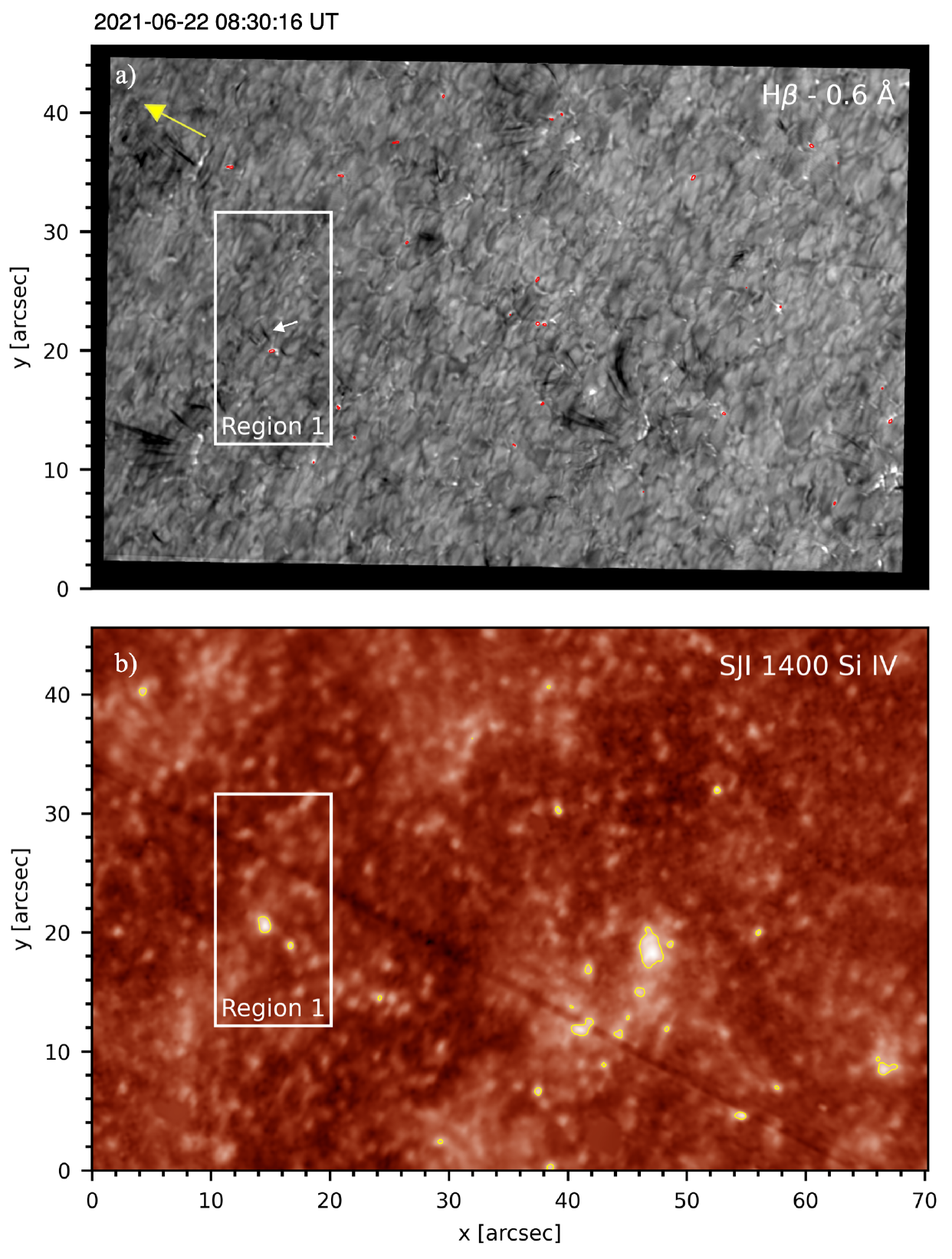}
    \caption{Overview of the observed region in \Hbeta\ blue wing (top) and SJI~1400 (bottom). The white rectangle marks the region used for studying the QSEBs and UV brightenings. Red contours in the top panel denote the QSEB detections. The dark elongated thread-like features are spicules. Inside Region~1, a spicule is visible close to the QSEB, shown by a white arrow, which later becomes part of an inverted-Y-shaped jet as discussed in Sect.~\ref{sec: 4.1}. Yellow contours in the bottom panel represent the detected $>$5$\sigma$ UV brightenings. The yellow arrow in the top panel shows the direction towards the north limb.}
    \label{fig:1}
\end{figure*}

\begin{figure*}[h!]
    \centering
    \includegraphics[width=0.9\linewidth]{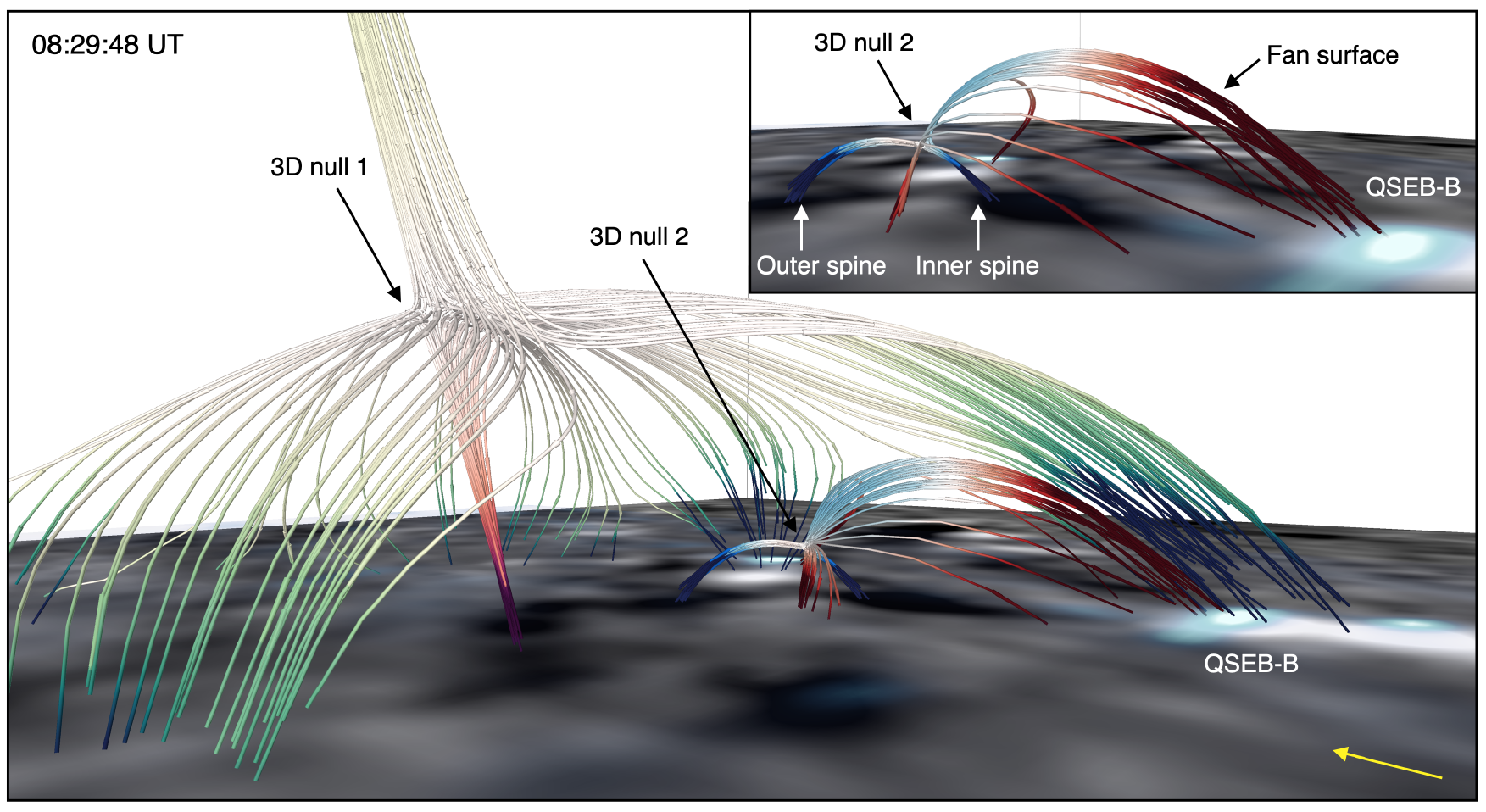}
    \caption{Two nested fan-spine magnetic field topologies obtained from potential field extrapolation. A smaller fan-spine structure (3D null 2) is located inside the fan surface of the larger one (3D null 1). The two fan surfaces share footpoints at two positive polarities, this is shown at a different angle in Fig.~\ref{fig:3}. QSEB-B is located at one of the shared footpoints of the fan-surfaces, shown in light blue and white shades in the \Hbeta\ $-0.6$~\AA\ image placed close to the photosphere. The inset shows the zoom-in of 3D~null~2 at a slightly different angle to clearly highlight the fan-spine configuration. The yellow arrow points to the direction of the north solar limb.}
    \label{fig:2}
\end{figure*}

\begin{figure*}[h!]
    \centering
    \includegraphics[width=\linewidth]{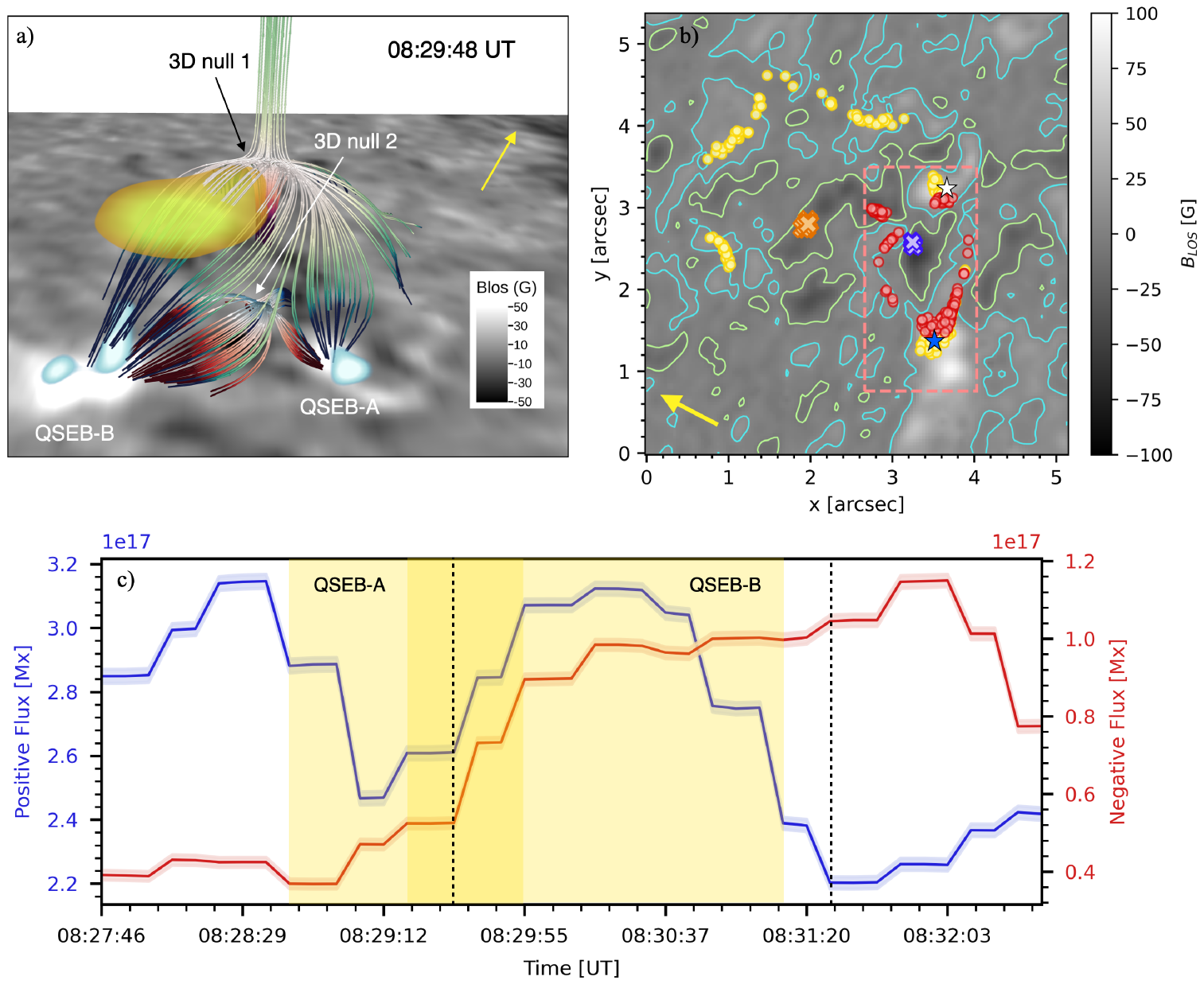}
    \caption{Magnetic field topology of a smaller fan-spine structure inside a bigger fan-spine structure. Panel a) shows the nested topology with the UV brightening in yellow color located close to the 3D null point associated with the bigger fan-spine topology. The spines of both fan-spine structures are rooted in different negative polarity patches. The QSEBs are shown in light blue and white shades in \Hbeta\ $-0.6$~\AA\ image placed close to the photosphere. Two QSEBs, namely QSEB-A and QSEB-B occur at two positive polarity patches, which are the shared footpoints of both the fan surfaces. Panel b) shows the $B_\mathrm{LOS}$ map with contours at 1$\sigma$ above the noise level with different markers: orange cross markers denote the inner spine footpoint of 3D null 1 and yellow circles denote the footpoints of its fan surface. The blue crosses denote the position of the inner spine footpoint of 3D null 2, and the red circles show the footpoints of its fan surface. The red circles on the shared negative polarity denote the outer spine footpoint of the 3D null 2 (around x, y = 3\arcsec, 3\arcsec). The location of QSEB-A is marked with a white star, while the location of QSEB-B is marked with a blue star. The pink dashed rectangle shows the area used for calculating the magnetic flux shown in panel c). The yellow arrows in the top panels show the direction towards the north limb. Panel c) shows the variation of positive and negative magnetic flux during the evolution of QSEBs and the inverted-Y-shaped jet. The error bands in positive and negative fluxes are shown as thin-shaded regions along the curves. The yellow-shaded regions denote the period of occurrence of QSEB-A and QSEB-B. The vertical dashed lines denote the start and end of the inverted-Y-shaped jet.}
    \label{fig:3}
\end{figure*}
UV bursts occur in the upper solar atmosphere and are found in regions with underlying opposite magnetic polarities. They are observed as compact, intense, and rapidly varying brightenings in the \Siiv\ spectral lines \citep{2014Sci...346C.315P, 2018SSRv..214..120Y}, in observations from the Interface Region Imaging Spectrograph \citep[IRIS,][]{2014SoPh..289.2733D}. 

They are characterised by broad \Siiv\ emission lines and include absorption blends from neutral or singly ionised species such as \FeII\ and \NiII. 
These spectral signatures indicate that plasma with transition region temperatures (100,000 K) is embedded beneath a cooler chromospheric canopy of fibrils \citep{2014Sci...346C.315P}. 
This has been corroborated by various follow-up studies \citep[e.g.,][]{2015ApJ...812...11V, 2015ApJ...809...82G, 2017MNRAS.464.1753H, 2022A&A...657A.132K}.
Using the IRIS observations of the \Mgtriplet\ and \Siiv\ lines, UV bursts have been found in close proximity to EBs by \citet{2015ApJ...812...11V}, who show that the tops of the EBs could heat plasma to transition region temperatures.
This is also supported by \citet{2016ApJ...824...96T} who reported 10 UV bursts associated with EBs. 
The first transition region response to a QSEB was reported by \citet{2017ApJ...845...16N} using \Halpha\ and IRIS \Siiv\ observations.
\citet{2019A&A...626A..33H} used 3D magnetohydrodynamic Bifrost simulations \citep{2011A&A...531A.154G} to show that EBs form in the lower photosphere (up to 1200~km), while UV bursts form at higher altitudes (700~km to 3~Mm), along extended current sheets as part of the same magnetic reconnection system. 
They further suggest that spatial offsets between EBs and UV bursts could be either due to the orientation of the current sheet or the viewing angle.
The observations of \citet{2015ApJ...812...11V}, \citet{2019ApJ...875L..30C}, and \citet{2020A&A...633A..58O} support that UV bursts appear with some offset toward the limb relative to EBs.
In the quiet Sun, weaker events compared to active region UV bursts, which are referred to as UV brightenings, have been observed in close association with QSEBs. 
In our previous study, \citet[][hereafter \citetalias{2024A&A...689A.156B}]{2024A&A...689A.156B}, we investigated the spatial and temporal relationship between QSEBs and UV brightenings.
We found that 15\% of long-lived QSEBs ($>1$~min) were associated with UV brightenings in the \Siiv\ lines, which typically occurred within 1000~km of the QSEB, often toward the solar limb. 
QSEBs also tend to occur before the UV brightenings. 
Some QSEBs were sampled by the IRIS slit and showed emissions in the \Siiv\ and \Mgtriplet\ spectral lines, indicating that they locally heat plasma to transition region temperatures.

Several scenarios based on magnetic topology have been suggested for the formation of UV bursts, such as, at bald patches in emerging flux regions \citep{2017ApJ...836...63T, 2017ApJ...836...52Z}, or in regions with high squashing factor \citep{1996A&A...308..643D, 2002JGRA..107.1164T, 2005LRSP....2....7L} approximately 1~Mm above the surface \citep{2018ApJ...854..174T}. 
They can also be triggered due to magnetic reconnection between newly emerging magnetic domes and the pre-existing ambient magnetic field, which can lead to the formation of a three-dimensional (3D) magnetic null as demonstrated by \citet{2017ApJ...851L...6R} and \citet{2017ApJ...850..153N}.
A 3D magnetic null point has a characteristic fan-spine topology which is made up of a dome-shaped fan surface and spine field line which meet at the null point where the magnetic field strength vanishes \citep{2002A&ARv..10..313P, 2005LRSP....2....7L}. 
The inner and outer spines extend through this null point, with their footpoints rooted in regions of the same magnetic polarity. Meanwhile, the fan surface anchors to a ring-shaped area of opposite polarity surrounding the inner spine.
This null point serves as a prime site for magnetic reconnection, releasing energy and triggering UV bursts, as has been shown in works by \citet{2017A&A...605A..49C} and \citet{2018A&A...617A.128S}.

The high-quality observations of QSEBs and UV brightenings in \citetalias{2024A&A...689A.156B}, which include detailed photospheric magnetic field measurements, formed the foundation for this study.
In \citet[][hereafter \citetalias{2025A&A...693A.221B}]{2025A&A...693A.221B}, we used the magnetic field data to perform potential field extrapolations and identified four magnetic topologies that can explain the formation of QSEBs and UV brightenings. These topologies included simple dipole and complex fan-spine configurations with 3D null points. The study provided observational support for each of the topological scenarios. For the cases involving the 3D null, UV brightenings were found to occur near the null point, while QSEBs were located at the footpoints of the inner spine, outer spine, and fan surface.
In this paper, we delve into more complex topological configurations involving interactions between two fan surfaces, which we associate, not only with QSEBs and UV brightenings but also with the formation of chromospheric inverted-Y-shaped jets.

\begin{figure}[h!]
    \centering
    \includegraphics[width=0.95\linewidth]{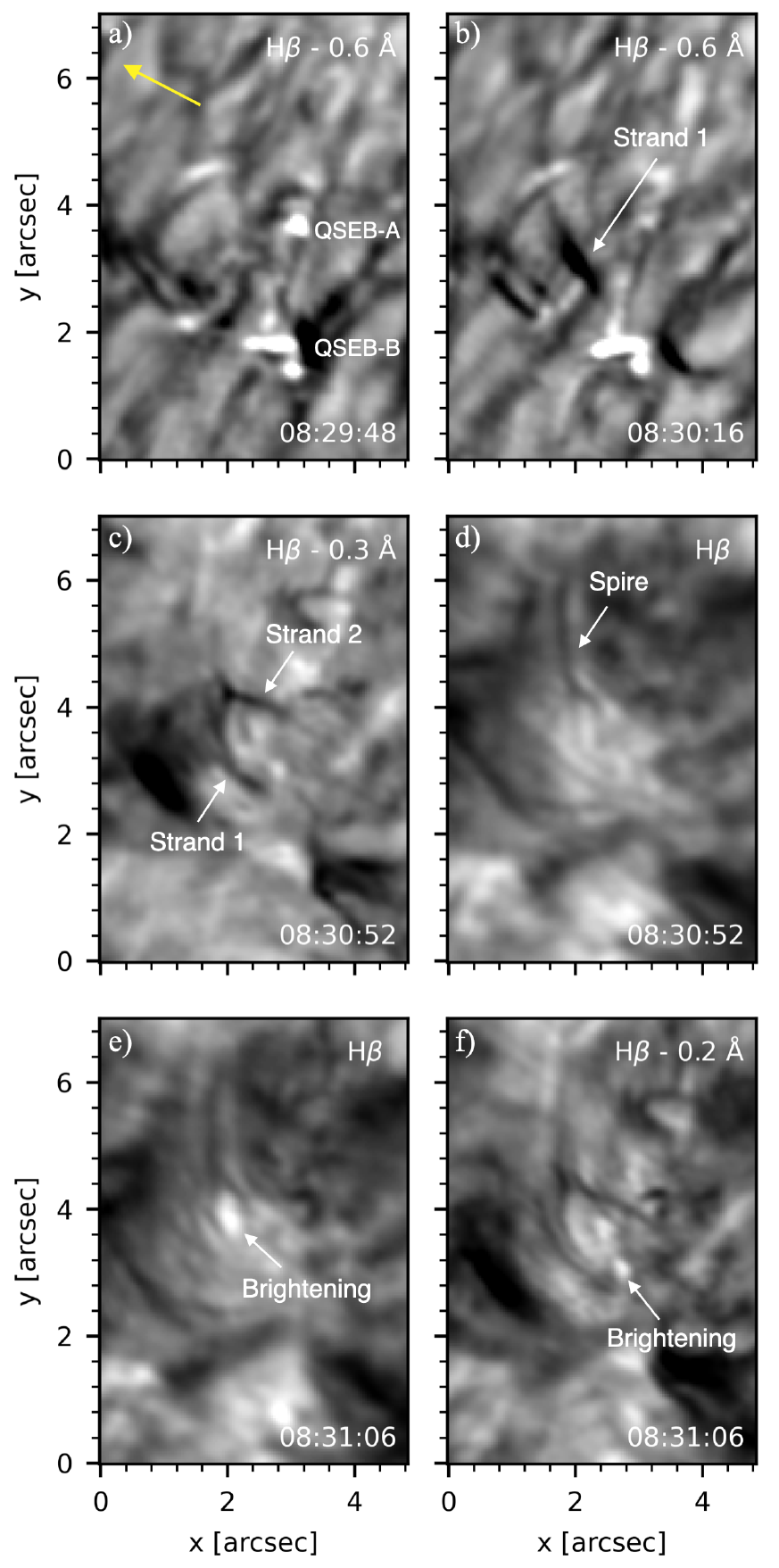}
    \caption{Evolution of the inverted-Y-shaped jet with QSEBs at its footpoints. Panel a) shows the two QSEBs in the \Hbeta\ wing. The yellow arrow points to the limb direction. Panel b shows the footpoints of the inverted-Y-shaped structure marked as Strand~1 and 2. Strand~1 originates close to QSEB-B and is shown in panels b) and c). Strand~2 starts later from the location of QSEB-A and is shown in panel c) when it merges with Strand~1. Panel d) shows the spire of the jet-like top part of the inverted-Y-shaped jet. Panel e) points to the brightening in the core of the \Hbeta\ line, which occurs just below the point of intersection of the two strands, while panel f) points to a brightening at the base of this structure. Note that the wavelength positions are different in the panels, to best show the inverted-Y-shaped jet at different instances. The QSEBs are shown only in panels a) and b) due to their clear visibility in the wing positions, and they fade before the jet itself dissipates. They are marked in panel a). The full evolution of the inverted-Y-shaped jet can be viewed in the accompanying movie (see \url{http://tsih3.uio.no/lapalma/subl/qseb_uvb_jet_topology/fig04.mp4}).}
    \label{fig:4}
\end{figure}
\section{Observations}
\label{sec:observations}

We analyzed the same observations that were used in \citetalias{2024A&A...689A.156B} 
and \citetalias{2025A&A...693A.221B}. 
A coronal hole near the North limb ($\mu=0.48$) was observed on 22 June 2021 for 51~min starting from 08:17:52~UT. 
The observations were part of a coordinated observation campaign between the SST and IRIS 
\citep{2020A&A...641A.146R}. 
From the SST, we used \Hbeta\ spectral line scans acquired with the CHROMIS instrument \citep{Scharmer2017} 
at a temporal cadence of 7~s. 
Furthermore, we used line-of-sight magnetic field maps ($B_\mathrm{LOS}$) derived from Milne-Eddington inversions 
\cite[using the code by][]{2019A&A...631A.153D} 
of spectro-polarimetric data in the \ion{Fe}{i} 6173~\AA\ line acquired with the CRISP instrument 
\citep{2008ApJ...689L..69S} 
at a cadence of 19~s.
The observations were processed following the SSTRED data reduction pipeline
\citep{2015A&A...573A..40D, 2021A&A...653A..68L}
which includes multi-object multi-frame blind deconvolution
\citep[MOMFBD,][]{2005SoPh..228..191V} image restoration.
The observations further benefited from the SST adaptive optics system \citep{2024A&A...685A..32S}. 
From IRIS, we used slit jaw images (SJI) in the 1400~\AA\ channel that is dominated by emission in the transition region \Siiv~1394~\AA\ and 1403~\AA\ spectral lines.
The cadence of the SJI images was 18~s. 
For more details on the observations and alignment between the different spectral lines and channels, we refer to \citetalias{2024A&A...689A.156B}. 

\begin{figure*}[h!]
    \centering
    \includegraphics[width=0.98\linewidth]{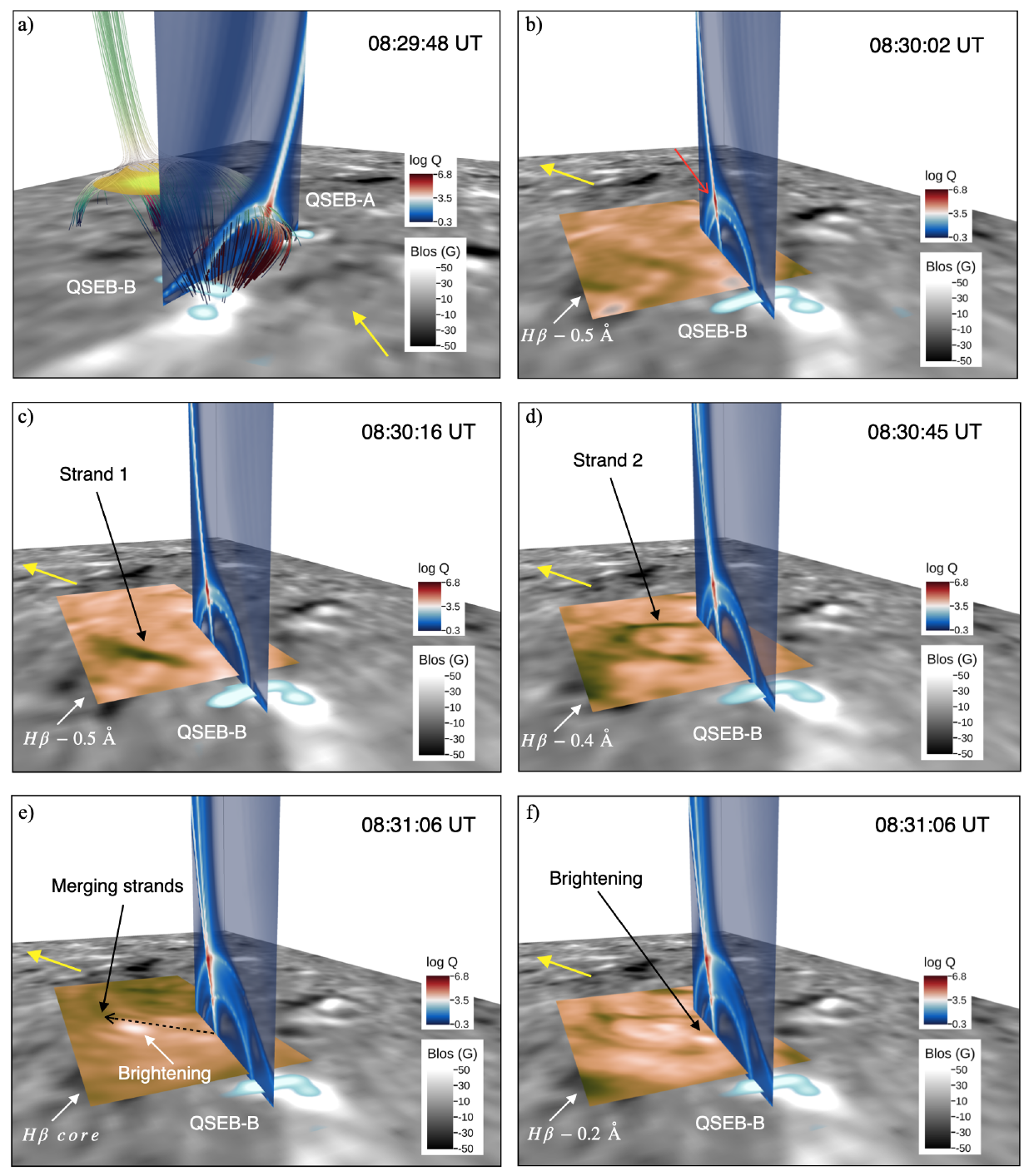}
    \caption{Stages of the inverted-Y-shaped jet along with the logarithm of the squashing factor ($\log~Q$), at different instances. In all panels, the grey-scale image at the bottom is the $B_\mathrm{LOS}$ map, while the yellow arrow points to the north limb. Panel (a) shows the magnetic field lines associated with the two 3D nulls, along with the UV brightening in yellow close to the outer 3D null and the QSEBs at the fan surface footpoints. The magnetic field lines are not shown in other panels to avoid clutter. All panels include $\log~Q$ slices, where red indicates high $Q$ values. The red arrow in panel (b) points to the region with the highest value along a vertical high $Q$ line, where reconnection likely occurs. A brown layer in panels (b) to (f) shows the \Hbeta\ image at different wavelengths, which depict the various features of the inverted-Y-shaped jet. The black dashed arrow in panel (e) marks the distance between the vertical high $Q$ line and the merging strands (1.2~Mm).}
    \label{fig:5}
\end{figure*}
\section{Method of analysis}
\label{sec:methods}

\begin{figure*}[h!]
    \centering
    \includegraphics[width=\linewidth]{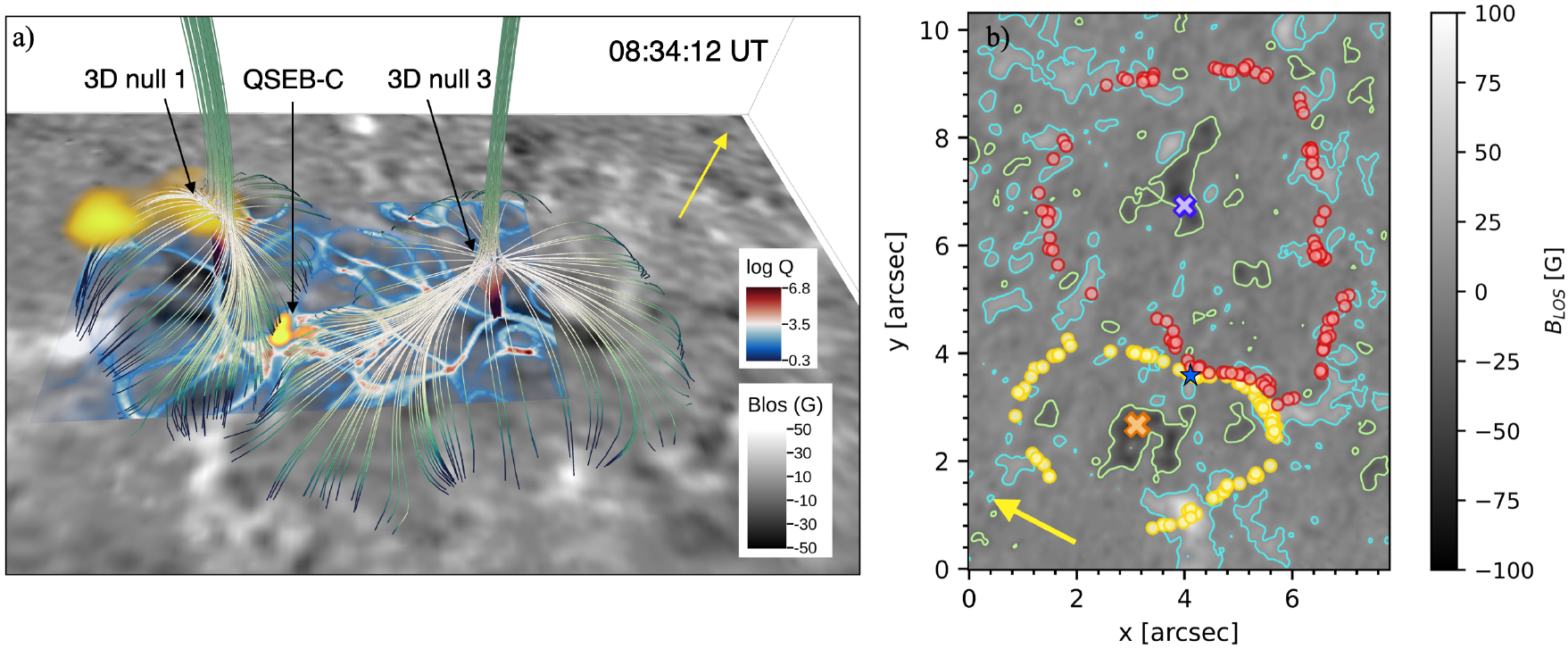}
    \caption{Magnetic field topology showing two adjacent fan-spine structures. The 3D~null~1 shown here is the same as in Fig.~\ref{fig:3}. The UV brightening is located close to the 3D~null~1. The inner spines of the two 3D nulls are rooted in different negative polarity patches, with their fan footpoints at nearby positive polarities. The QSEB is shown in yellow colour in \Hbeta\ $-0.6$~\AA\ image for better visibility among a large number of magnetic field lines. QSEB-C is located at the shared fan surface footpoints of both the 3D nulls. Panel b) shows the $B_\mathrm{LOS}$ map with contours at 2$\sigma$ above the noise level with different markers: orange cross markers denote the inner spine footpoint of the 3D~null~1 and yellow circles denote the footpoints of its fan surface. The blue crosses denote the position of the inner spine footpoint of the 3D~null~3, and the red circles show the footpoints of its fan surface. The location of QSEB-C is marked with a blue star. The $\log~Q$ plane is displayed close to the photosphere, where regions in red denote high values of $Q$. Yellow arrows in both panels show the direction towards the limb.}
    \label{fig:6}
\end{figure*}
\subsection{Identification of events}
A complete explanation of the methodology for identifying QSEB events in the \Hbeta\ data, the corresponding UV brightenings in the SJI~1400 data, and the procedures for linking them is provided in \citetalias{2024A&A...689A.156B}. 
To summarise, our QSEB detection method uses \textit{k}-means clustering \citep{Everitt_1972} to find characteristic EB spectral profiles and employs connected component analysis to link them spatially and temporally.
Each QSEB event is tracked using the Trackpy Python library\footnote{\url{https://soft-matter.github.io/trackpy/v0.6.4/}} and is given an event ID number.
%
The study detected 1423 QSEB events during the 51-min observation period. 
A threshold of 5$\sigma$ above the median background was applied to detect the brightest UV events, resulting in 1978 detections. 
Many of the associated events were found to be within 1000~km of the QSEBs.
\citetalias{2025A&A...693A.221B} studied two regions of the same dataset, namely Region 1 and Region 2, where multiple QSEBs and UV brightenings were observed close to each other.
In this study, we focus on events within Region 1, but with larger dimensions to better capture the complex topological structures (see Fig.~\ref{fig:1}). 
This region features recurring QSEBs and UV brightenings, along with chromospheric inverted-Y-shaped jets in \Hbeta.
Figure~\ref{fig:1} highlights an example where a QSEB is visible in the \Hbeta\ wing, accompanied by a nearby $>$5$\sigma$ UV brightening detected in SJI~1400 within Region~1.

\begin{figure}[h!]
    \centering
    \includegraphics[width=\linewidth]{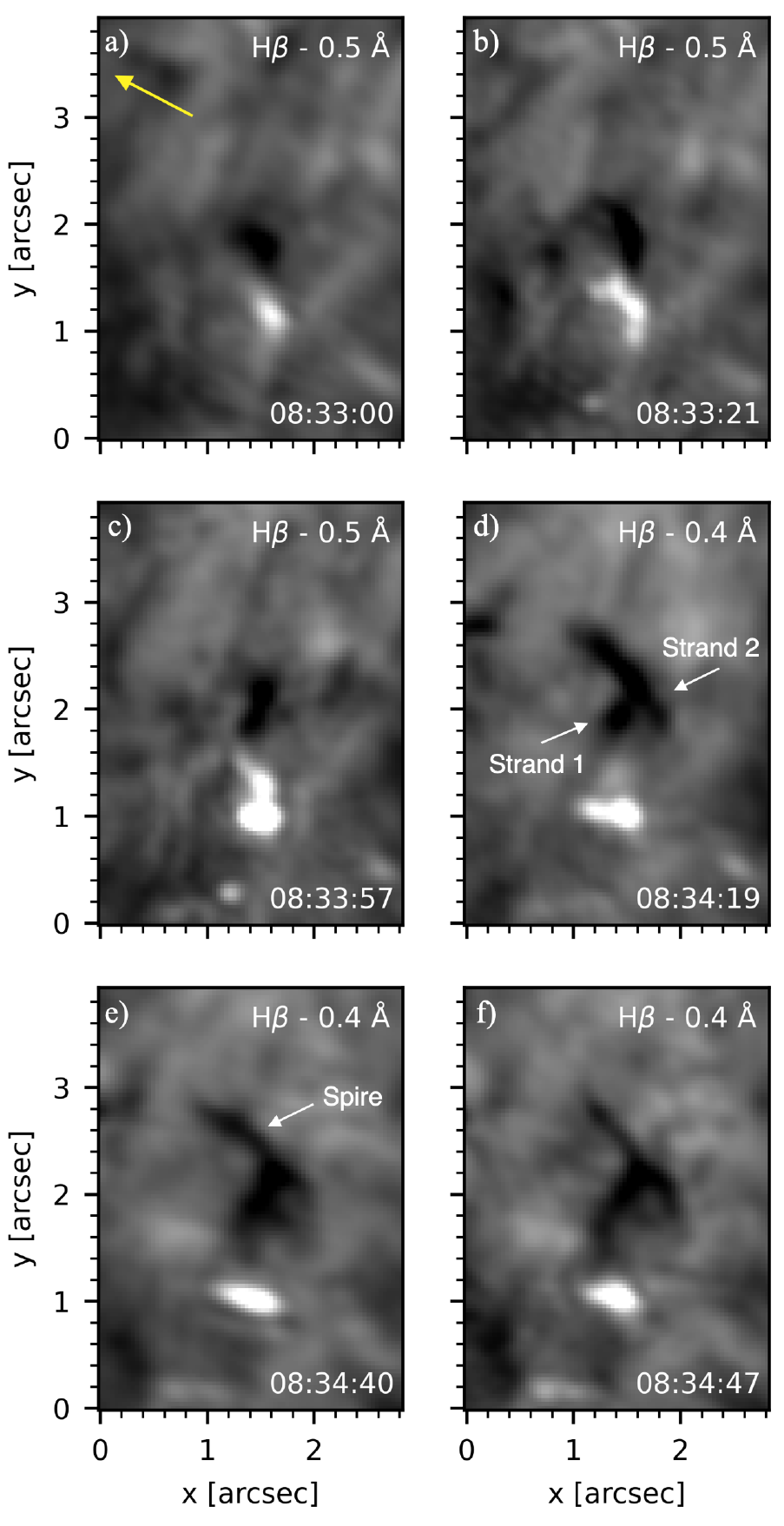}
    \caption{Evolution of the inverted-Y-shaped jet alongwith QSEB-C of Fig.~\ref{fig:6} at one of its footpoints. The footpoints of the inverted-Y-shaped structure are marked as Strand 1 and 2. Panel a) shows the QSEB in the \Hbeta\ wing, along with the beginning of Strand 1 of the inverted-Y-shaped jet, which originates very close to the QSEB. Both the QSEB and Strand 1 begin at the same time. Panels b) and c) show the progress of QSEB and the strand. Strand 2 starts to develop in panel c) but is clearly visible in panel d), e) and f). The merged strands with the spire of the inverted-Y-shaped jet are visible in panels d), e) and f). The yellow arrow in panel a) points to the limb direction. Note that the wavelength positions are different in the panels, to best show the inverted-Y-shaped jet at different instances. The full evolution of the inverted-Y-shaped jet can be viewed in the accompanying movie (see \url{http://tsih3.uio.no/lapalma/subl/qseb_uvb_jet_topology/fig07.mp4}).}
    \label{fig:7}
\end{figure}

\subsection{Magnetic field extrapolation}
To infer the magnetic field topology associated with the QSEB events, we applied Fast Fourier Transformation (FFT) based potential field extrapolations \citep{1972SoPh...25..127N, 1981A&A...100..197A} on Region~1 using the photospheric $B_\mathrm{LOS}$ data from SST. The extrapolation is performed over a box size of 256 $\times$ 512 $\times$ 256 grid points, approximately corresponding to a physical domain size of 7~Mm $\times$ 14~Mm $\times$ 7~Mm in the $x$, $y$, and $z$ directions, respectively. The bottom boundary for extrapolation was selected to ensure flux balance, allowing the resultant extrapolated magnetic field to closely satisfy the divergence-free condition. The mean value of $B_\mathrm{LOS}$ for this region is 0.66~G, which is much lower than the noise level of 6.4~G in $B_\mathrm{LOS}$. The ratio of the total flux to the total unsigned flux in this case is 0.09. 
The extrapolated magnetic field lines were visualised in 3D using the VAPOR software \citep{2019Atmos..10..488L}.
The VAPOR software allows us to trace the magnetic field lines near the base of the QSEB through bidirectional field line integration by randomly placing the seed points with a bias towards higher values of $|B_\mathrm{LOS}|$. This allows us to continuously follow the strong polarities in a small region close to the QSEB as they move during the time series. We further calculated the squashing factor $Q$ \citep{2002JGRA..107.1164T, 2009ApJ...693.1029T} for the extrapolated magnetic field using the code by \citet{2016ApJ...818..148L}. This allowed us to locate the magnetic null points by biasing the seed points with a large squashing factor, and follow the changes in fan-spine topology associated with the 3D null points during the event. To visually compare the 3D magnetic field lines with QSEBs in the \Hbeta\ wing and UV brightenings in the SJI~1400~\AA\ channel, we have placed the \Hbeta\ layer in a plane close to the photosphere, while the SJI~1400 layer is placed at different heights depending on the 3D null's location.

\section{Results}
\label{sec:results}
This study presents two scenarios where QSEBs and associated UV brightenings arise due to interaction between the fan surfaces of two 3D nulls.
In the first case, the fan surface of one 3D null resides inside a larger fan surface of another 3D null. In the second case, the two fan surfaces are situated side by side, with a small region of overlap. 
In both cases, we observe chromospheric inverted-Y-shaped jets originating close to the QSEBs where the fan surfaces of the two 3D nulls interact. 
In the following subsections, we present observations and magnetic field extrapolations illustrating these cases.

\subsection{Nested fan-spine topologies}  \label{sec: 4.1}

Here, we study two QSEBs events associated with two nested 3D nulls with a UV brightening observed close to the outer 3D null.
The potential field extrapolation for this region is shown in Fig.~\ref{fig:2}, which illustrates the nested fan-spine topologies. The inner fan-spine configuration (3D~null~2) is located inside an outer fan-spine configuration (3D~null~1). This configuration is also shown at a different viewing angle, at the same instance in Fig.~\ref{fig:3}a. The fan surfaces of both the 3D nulls are clustered in regions of local magnetic field concentrations coinciding with the location of the two QSEBs.
These two QSEBs, designated as QSEB-A and QSEB-B, are visible in Fig.~\ref{fig:3}a, which depicts an instance when both QSEBs occur simultaneously.
QSEB-A starts at 08:28:43 UT and ends at 08:29:55 UT, lasting for a duration of 72~s, while QSEB-B begins at 08:29:12 UT and ends at 08:31:13 UT with a duration of 121~s. 
Notably, the UV brightening occurs close to the outer 3D null point and is situated 660~km above the photosphere in Fig.~\ref{fig:3}a.
This UV brightening is the same as the one previously discussed in section 4.3 of \citetalias{2025A&A...693A.221B}. 
%
For the sake of completeness, here we briefly summarise this scenario from section 4.3 of \citetalias{2025A&A...693A.221B} that included an observation of a UV brightening occurring close to the 3D null 1, and the QSEB being situated close to the footpoints of the fan surface having a local concentration of strong magnetic field. The QSEB was also associated with a dipolar flux-concentration whose positive footpoint was located on the same polarity as that of the fan surface footpoints. 
There, the QSEB and UV brightening were likely caused by a common reconnection process due to the formation of a quasi-separatrix layer (QSL) between the emerging dipole and the fan surface. 
In the present study, we continue following the evolution of this same region. The event occurs approximately 3~min after the one described in section 4.3 of \citetalias{2025A&A...693A.221B}, where now instead of an emerging dipole we have an emerging 3D null with fan-spine topology. We now observe two QSEBs: QSEB-A and QSEB-B, at the footpoints of the initial fan-spine configuration. The UV brightening continues from the previous event and likely occurs due to the formation of QSLs between the emerging fan surface and the already existing outer fan surface.

%
%
%
Figure~\ref{fig:3}b marks the footpoints of field lines close to the inner spines and fan surfaces of the two 3D null points, as well as the locations of the QSEBs on the $B_\mathrm{LOS}$ map.
The pink dashed box outlines a region containing the two stronger positive polarity patches, where the QSEBs are located. 
%
%
Figure~\ref{fig:3}c shows the evolution of positive and negative magnetic flux in this region. 
We notice that an episode of flux emergence starts around 08:29:05 UT. The positive flux then begins to decrease from 08:30:37 UT, although the negative flux continues to increase, suggesting possible cancellation along with flux emergence.

%
The two dashed vertical lines in Fig.~\ref{fig:3}c mark the start and end of a chromospheric inverted-Y-shaped jet that originates close to the two QSEBs. 
Figure~\ref{fig:4} illustrates different stages of this inverted-Y-shaped jet in different wavelength positions of the \Hbeta\ line at different times. During the evolution, we observe the presence of two strands and a spire of chromospheric material. 
%
%
Notably, the jet is visible in the blue wing of the \Hbeta\ spectral line, implying that these are related to possible reconnection outflows, or similar to RBEs.
The inverted Y-shaped jet originates as a single strand (Strand 1) at 08:29:33 UT and is visible in Fig.~\ref{fig:4}b, which displays the blue wing image at 08:30:16 UT.
Figure~\ref{fig:1} also depicts the \Hbeta\ FOV at this time, in which the QSEB and the strand are visible inside the white rectangle. The strand looks quite similar to the other spicules (RBEs), which are dark, elongated, thread-like structures in the image. 
This strand originates close to QSEB-B, at coordinates ($x,y$ = 3\farcs2, 1\farcs8) as shown in Fig.~\ref{fig:3}b.
From the online movie, it can be seen that this strand bends at its top around 08:30:30 UT.
Another strand (Strand 2) appears slightly later, at 08:30:23 UT, from the location of QSEB-A.
Both of the strands originate during the flux emergence episode shown in Fig.~\ref{fig:3}c.
The two strands meet at 08:30:30 UT, which is shown in panel (c) at 08:30:52 UT. 
Figure~\ref{fig:4}d shows the spire of the jet in the core of the \Hbeta\ line at 08:30:52 UT. 
This spire is visible in the \Hbeta\ core from 08:29:48 UT. 
Figure~\ref{fig:4}e shows the two strands converged at the base of the spire at 08:31:06 UT, which then resembles the inverted-Y-shaped jet.
After the strands meet, we also observe some brightening in the \Hbeta\ core just below their point of intersection. 
This brightening is shown in Fig.~\ref{fig:4}e at 08:31:06 UT, and lasts for 28~s from 08:30:52 UT to 08:31:20 UT. 
As the jet rises upwards in the \Hbeta\ core, the brightening below their point of intersection also moves up with the jet. 
The jet fades after 08:31:28 UT across all the wavelength positions of the \Hbeta\ spectral line.
Additionally, another brief brightening is detected near the midpoint of the two QSEBs, which is shown in Fig.~\ref{fig:4}f in the \Hbeta\ $-0.2$~\AA\ image. 
This brightening is short-lived and is visible from 08:30:59 UT to 08:31:06 UT.
The full event, starting from the beginning of Strand 1, the merging of strands, then the brightening in the \Hbeta\ core, brightening at the footpoint to the disappearance of the inverted-Y-shaped jet, lasts for 115~s and can be seen in the corresponding online movie. 
%
%

In the absence of electric currents in the potential field extrapolation, we use the squashing factor $Q$ to relate the evolution of the inverted-Y-shaped jet with the magnetic topology of the region.
%
%
Figure~\ref{fig:5} shows the different stages of the inverted-Y-shaped jet alongside the logarithm of the squashing factor ($\log~Q$). 
Figure~\ref{fig:5}a corresponds to the same instance depicted in Fig.~\ref{fig:3}a and Fig.~\ref{fig:4}a. 
%
%
The magnetic field lines of the two fan surfaces converge at the same two positive polarity patches where the QSEBs A and B occur. 
The flux emergence along with the convective motions in the solar photosphere can cause a potential misalignment between the newly-emerged inner fan-spine structure (3D null 2) and the pre-existing outer fan-spine structure (3D null 1), which could lead to the formation of QSLs and current sheets, leading to subsequent magnetic reconnections. This could be the reason for UV brightening occurring close to the 3D~null~1.
%
In the panels, we show a slice of the squashing factor $Q$ in a plane connecting the two QSEBs, which passes through the two fan surfaces. 
The high $Q$ values in this plane show a dome and spine-like contour with the footpoints of this contour connecting the positive polarity patches where the two QSEBs occur. 
The region with the highest $Q$ value along a vertical high $Q$ line in this plane is indicated with a red arrow in panel (b), where reconnection can potentially take place. 
The spire of the inverted-Y-shaped jet could be due to an outflow after the reconnection at this high $Q$ region, and lie along the outer spine, which is aligned with the direction of the open magnetic field lines. 
Panels (b) and (c) show two instances with Strand 1 of the jet and the QSEB-B in the wings of the \Hbeta\ line. 
The reconnection at QSLs could trigger both the QSEB-B and Strand 1 of the jet, with QSEB occurring at the footpoint and the strand originating close to the QSEB that follows the path along the high $Q$ contour. 
Panel (d) depicts Strand 2 of the jet after it merges with Strand 1 in the wings of the \Hbeta\ line. 
Since this strand originates from the site where QSEB-A was previously located, the cause of its occurrence is probably similar to that of Strand 1. 
Notably, the contour of high $Q$ values close to QSEB-B is more extended than the one close to QSEB-A. 
This may explain why Strand 1 is longer compared to Strand 2 as seen in Fig.~\ref{fig:4}c.
The brightening in the \Hbeta\ core just below the point of intersection of the two strands likely occurs below the reconnection site, along the high $Q$ region, as seen in panel (e). 
%
%
Figure~\ref{fig:5}f points to a brightening close to the footpoint of the vertical high $Q$ line. This footpoint coincides with the inner spine footpoint of the 3D~null~2 (field lines are not shown in this panel).
%
Since our observation is close to the north solar limb, we see the jet projected towards the direction of the limb in the \Hbeta\ images.
Projection effects corresponding to a viewing angle of $\mu=0.48$ lead to an offset of approximately 1.78~Mm in the observed position for every 1~Mm of height in the solar atmosphere.
The high $Q$ region (shown in panel (b) with a red arrow) is situated between 660~km to 720~km from the footpoint of the inner spine. 
Since the two strands likely intersect at the region of high $Q$ values, this translates to a distance of 1.1~Mm to 1.2~Mm due to the projection effects.
This matches with the distance of the point of intersection of the two strands in \Hbeta\ core from the footpoint of the vertical high $Q$ line is 1.2~Mm. 
From the observations, we also see that the jet rises upwards in the \Hbeta\ core and fades by 08:31:28 UT.
This agrees with the instance when the inner 3D~null~2 is no longer present in the magnetic field extrapolations (not shown here). 
From panels (a) to (f) of Fig.~\ref{fig:5}, we note that the size of the $Q$ contour keeps on increasing from 08:29:48 UT to 08:31:06 UT. 
The region with the highest $Q$ values is around 386~km above the photosphere at 08:29:48 UT and rises to 720~km by 08:31:06 UT. 
%
The increase in the magnetic field strength at the negative polarity during flux emergence can explain the upward expansion of the high $Q$ region, while the subsequent magnetic reconnection can explain the rising nature of the chromospheric inverted-Y-shaped jet.

\subsection{Adjacent fan-spine topologies}  
\label{sec: 4.2}

In this section, we present a scenario involving two fan-spine structures corresponding to two null points (3D nulls 1 and 3) that are adjacent to one another. A QSEB (QSEB-C) is located at a shared polarity patch where both the fan surface footpoints are situated.
%
This magnetic field configuration is shown in Fig.~\ref{fig:6}a.
About 3~min after QSEB-A, the positive polarity associated with it in Sect.~\ref{sec: 4.1} moves slightly towards the north, likely due to convection, and QSEB-C is later observed at this polarity. 
QSEB-C is still located at the footpoints of the 3D~null~1 discussed in the previous section, although the inner fan-spine structure (3D~null~2) does not exist any more.
QSEB-C is highlighted with a different colour as compared to previous figures for better visibility among the large number of magnetic field lines shown in the panel.
We also observe a UV brightening close to the 3D~null~1 on the left, which has been persistent since the start of QSEB-A and is highlighted in yellow near the null 1 in panel (a). The height of the 3D~null~1 has increased to 860~km since the previous event, while the 3D~null~3 is located 1012~km above the photosphere.
Panel (b) highlights the footpoints of the two fan surfaces, along with the position of the footpoints of their inner spines, and the QSEB marker at the intersection of the fan surfaces.
In panel (a), we have shown the logarithm of the squashing factor $\log~Q$ at a height close to the photosphere. 
We notice a region of high $Q$ (in red), below the QSEB, above the shared polarity of the two fan-spine topologies.
The high $Q$ region persists for the whole duration of the QSEB, suggesting the probable formation of the QSLs at the intersection of fan surfaces, which likely causes current sheets formation and drives the QSEB activity.
For this case, it was difficult to calculate the magnetic flux as the two fan-spine structures cover a large area and involve many positive polarities, which are not associated with the QSEB.
We also observe the formation of a chromospheric inverted-Y-shaped jet, emerging from this QSEB. 
The various stages of this jet are illustrated in Fig.~\ref{fig:7}, which shows six instances at different times and wavelength positions in the wings of the \Hbeta\ line.
QSEB-C begins at 08:32:46 UT and at the same time, Strand 1 of the jet starts to appear close to the QSEB, depicted in Fig.\ref{fig:7}a.
Strand 2 forms at 08:33:57 UT, and shortly afterwards, the spire becomes visible, extending towards the direction of the limb, and completing the inverted-Y-shaped jet. In this case, we were not able to isolate the magnetic field structures associated with the strands, as it is likely that these structures form as a result of the interaction between the fan surfaces and are missing in the potential field extrapolation.
QSEB-C persists for 164~s, ending at 08:35:31 UT.
Unlike the previous case, the jet in this case appears to fall down and fades by 08:37:04 UT, lasting for 258~s.
An accompanying movie also shows the evolution of the QSEB and the jet.

\section{Discussion}
\label{sec:discussion}
This study investigates scenarios involving the interaction between two fan-spine topologies that are associated with QSEBs, UV brightenings, and chromospheric inverted-Y-shaped jets. 
The QSEBs are studied using the \Hbeta\ data from the SST/CHROMIS instrument, and are detected using the \textit{k}-means clustering algorithm. 
The UV brightenings are identified from the IRIS SJI~1400 data, using a threshold of 5$\sigma$ above the median. 
Potential field extrapolations were performed on the high-resolution magnetograms from SST to study the evolution of the magnetic field topology in these regions. 
Our observation region is a coronal hole close to the north limb, hence the magnetic field extrapolations are done using only the line of sight magnetic field, due to substantial noise in the transverse field components ($B_\mathrm{x}$ and $B_\mathrm{y}$) and projection effects. 
The limitations of the dataset and methods have been carefully discussed in detail in \citetalias{2025A&A...693A.221B} and are summarised in the following section.

\subsection{Limitations}
In this section, we briefly discuss the limitations of our methodology and their implications for the results presented in this study. Our analysis is based on observations of a quiet Sun region, which is characterised by relatively weak magnetic fields. This region is located near the northern limb and hence has significant projection effects when comparing features across different heights. Due to these reasons, the measurements of the local transverse magnetic field components ($B_x$ and $B_y$), which are derived from the Stokes $Q$ and $U$ profiles, were noisy. This made it difficult to resolve the $180^\circ$ ambiguity and accurately obtain the vector magnetic field. To mitigate these issues, we based our extrapolations solely on the line-of-sight magnetic field ($B_{\text{LOS}}$), which is derived from Stokes $V$, and performed a potential field extrapolation without correcting for projection effects. The assumption of absence of electric currents in potential field approximation is a reasonable approximation for quiet Sun regions (except near the reconnection sites), where weaker magnetic fields give rise to relatively small currents. Our analysis was based on identifying null points and analysing their connectivity. Although we understand that the location of a null point can vary depending on the extrapolation method and null-detection algorithm used, we believe that the fan-spine topology is quite robust, and for our purposes, only the overall structure is important and not the exact location. \cite{2009SoPh..254...51L} demonstrated that potential field extrapolations can accurately locate these structures.

\subsection{Discussion of the topologies}
As context, in \citetalias{2025A&A...693A.221B} we have shown, through a combination of observations and magnetic field modelling, that within a fan-spine configuration, a UV brightening can be observed close to the 3D null point and QSEBs can potentially be found at three different locations: (i) near the footpoint of the inner spine, (ii) near the footpoint of the outer spine, and (iii) near the footpoints of the fan surface. 
In Sect.~4.3 of \citetalias{2025A&A...693A.221B}, we presented an observation of a UV brightening occurring close to a 3D null point, and the QSEB happening close to the footpoints of the fan surface having a local concentration of strong magnetic field. 
%
\begin{figure*}[h!]
    \centering
    \includegraphics[width=\linewidth]{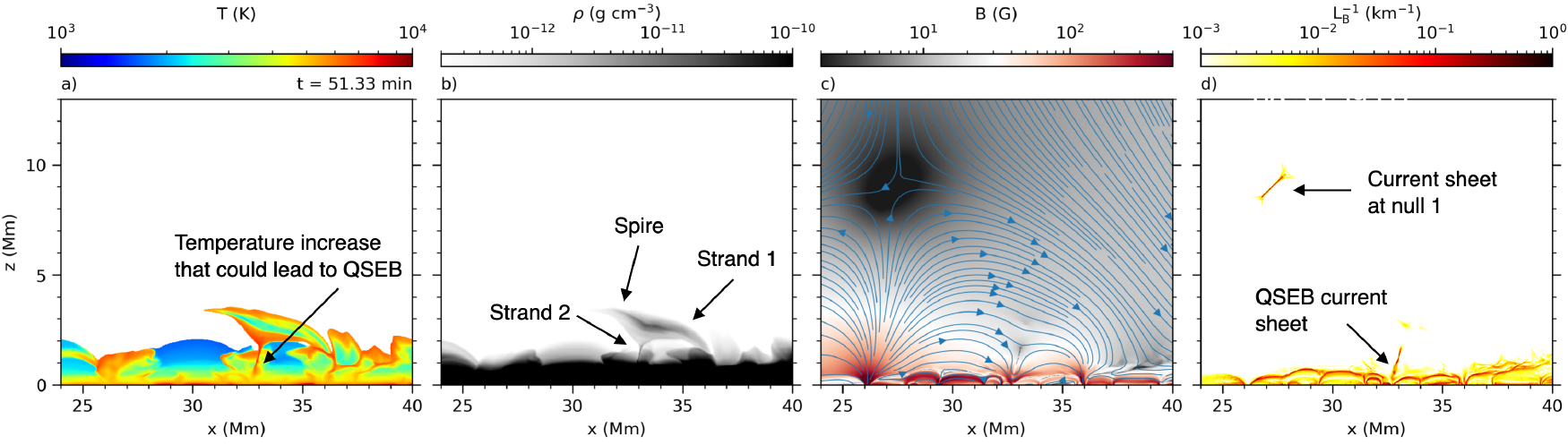}
    \caption{Chromospheric inverted Y-shaped jet from the numerical experiment by \cite{2022ApJ...935L..21N}.
    The panels show, from left to right: the temperature, T; the mass density, $\rho$; the magnetic field strength, B, with superimposed magnetic field lines; and the inverse of the characteristic length of the magnetic field, L$^{-1}_{\mathrm{B}}$.
    In panel a), the temperature is masked for densities smaller than $2\times10^{-13}$~g cm$^{-3}$ to ease the identification of chromospheric features.
    An animation of this figure is available online with the evolution of the system from $t=33.33$ to $t=53.30$~min (see \url{http://tsih3.uio.no/lapalma/subl/qseb_uvb_jet_topology/fig08.mp4}).
    }
    \label{fig:8}
\end{figure*}

We suggested that the QSEB and UV brightening were likely caused by a common reconnection process due to the formation of a QSL between the emerging dipole of the QSEB and the fan surface.
In this work, we revisit Region~1 from \citetalias{2025A&A...693A.221B} (now considering a larger FOV), which contains multiple recurrent QSEBs and associated UV brightenings that occur close to the 3D null point. The UV brightening discussed in Sect.~\ref{sec: 4.1} follows a similar scenario, but instead of an emerging dipole, we have an emerging 3D null with a fan surface. The same UV brightening persists (at the 3D null 1) for the event studied in Sect.~\ref{sec: 4.2}.

In Sect.~\ref{sec: 4.1}, we find that flux emergence leads to the formation of an inner-fan-spine topology inside the outer fan-spine topology. We also observe two QSEBs associated with this nested fan-spine configuration. The QSEBs are situated at the two shared polarities where the footpoints of the fan surfaces are located. 
A chromospheric inverted-Y-shaped jet also occurs, with the strands rooted close to the QSEBs. 
These small-scale events are likely driven by the formation of current sheets between the misaligned magnetic field lines of the two fan surfaces, leading to magnetic reconnection.  
We have found a similar scenario in the 2D numerical experiment by \cite{2022ApJ...935L..21N} using the Bifrost code \citep{2011A&A...531A.154G}. The simulation was initiated using a null configuration obtained from potential field extrapolation of a prescribed distribution at the bottom boundary. Though not directly based on the current observations, we refer to this experiment for illustration purposes to show that such scenarios of nested nulls can arise due to flux emergence following magnetic buoyancy instabilities like magnetic Rayleigh–Taylor instability \citep{1979SoPh...62...23A, 2014LRSP...11....3C} or the Parker instability \citep{1992ApJS...78..267N, 2004ApJ...614.1042M, 2007ASPC..369..355I}.
In this simulation, a small-scale flux emergence episode self-consistently took place inside the fan of a large fan-spine topology whose null point was located at coronal heights.
This event is illustrated in Fig.~\ref{fig:8} through maps of temperature, density, magnetic field, and characteristic length. The evolution of the magnetic field in panel Fig.~\ref{fig:8}c (also see accompanying movie) reveals the presence of a bald patch \citep{Titov_etal:1993} at $x$ = 33 Mm and $t$ = 43 min. Bald patches are topologically significant features, as they can lead to current sheet formation and, depending on the evolution, serve as precursors to null points with a fan-spine configuration \citep{2008AnGeo..26.2967M}. In the simulation, the bald patch collapses and evolves into a null point through plasmoid-mediated reconnection. Although this is not a one-to-one match with the observations, it highlights how reconnection can be triggered within a fan-spine structure following flux emergence. The characteristic length \citep{2016ApJ...822...18N}, which is defined as $\displaystyle{\mathrm{L}^{-1}_{\mathrm{B}} = \frac{\left| \nabla \times \vec{B} \right|}{|\vec{B}|}}$ facilitates the identification of the current sheet associated with the large fan spine and the one associated with the small-scale flux emergence beneath the large fan. In the latter, magnetic reconnection heats the chromospheric plasma, increasing the temperature by several hundreds of K, and launches a chromospheric inverted Y-shaped ejection. The vertical current sheet is located at $x$ = 32.5 Mm in Fig.~\ref{fig:8}d.
This event serves as a larger-scale version that resembles the observational scenario presented in Sect.~\ref{sec: 4.1}, where the reconnection could explain the QSEB in the lower atmosphere. The increased temperature that resulted from the reconnection could be sufficient to produce enhanced \Hbeta\ wing emission that would be observed as a QSEB. 
Figure~\ref{fig:8}b shows the strands and spire of the inverted-Y-shaped jet in the simulation, with Strand 2 originating along the current sheet that could lead to a QSEB. 
The accompanying movie highlights an additional current sheet near $x$ = 36.5 Mm up to $t$ =  44~min where another QSEB could be located, and where the Strand 1 seems to be rooted at $t$ = 51.33~min.

\begin{figure*}[h!]
    \centering
    \includegraphics[width=0.75\linewidth]{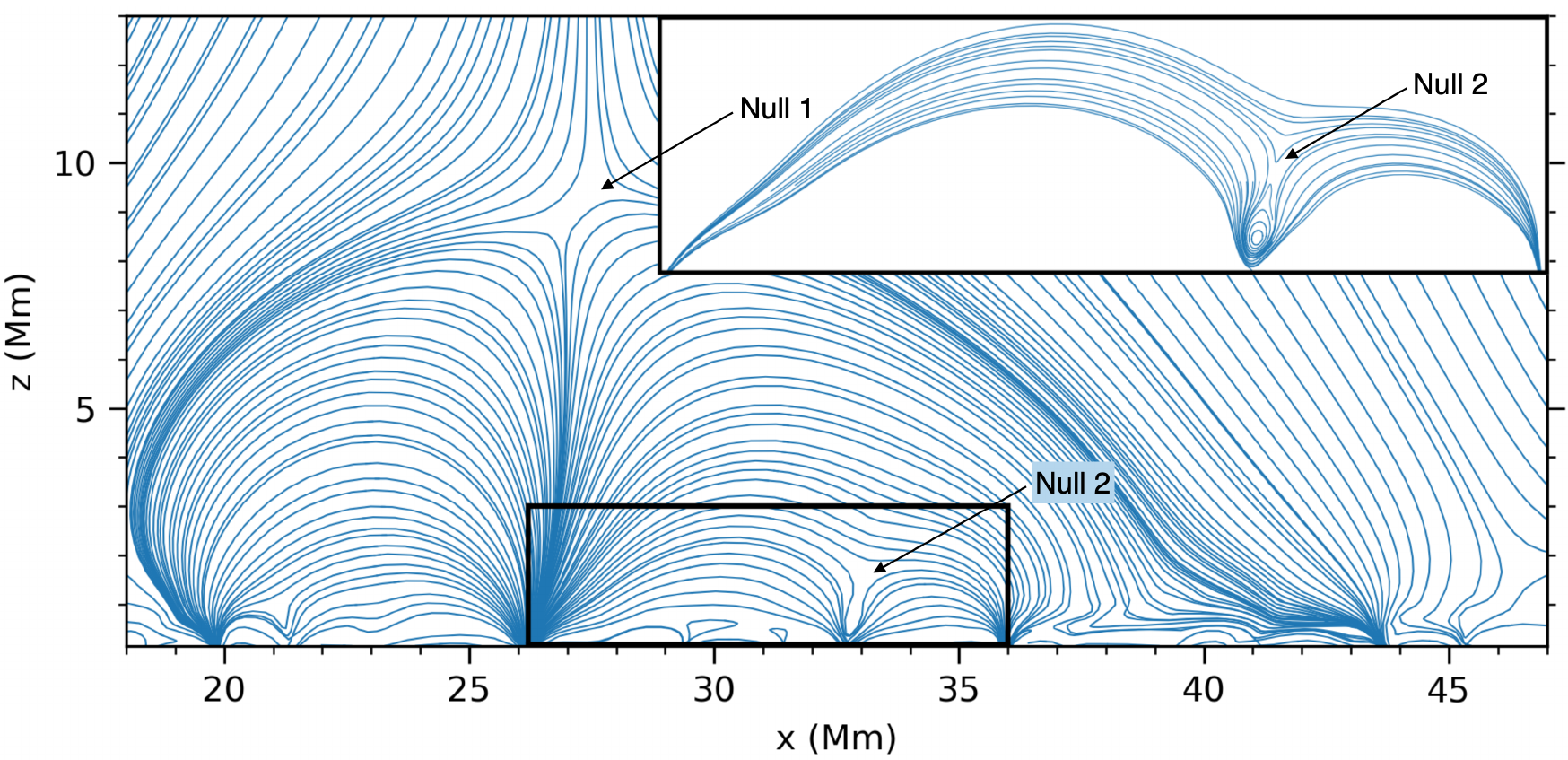}
    \caption{
    Detailed view of the magnetic field configuration from the 2D simulation at $t = 51.33$~min (corresponding to Fig.~\ref{fig:8}), showing the nested null configuration (with null 2 inside the fan of null 1).
    The rectangular inset shows a zoomed view of the magnetic structure around null 2, where only the relevant field lines are plotted for clarity.}
    \label{fig:9}
\end{figure*}

To further investigate the magnetic field configuration present in our 2D numerical simulation and its resemblance to the observations in greater detail, we show Fig.~\ref{fig:9}. This figure displays the magnetic field lines for the same simulation timestep as shown in Fig.~\ref{fig:8}, and highlights the nested null scenario similar to Fig.~\ref{fig:2}. Additionally, the rectangular inset zooms on the inner null where plasmoid-mediated magnetic reconnection could result in enhanced temperatures, causing the formation of a QSEB.
%

In Sect.~\ref{sec: 4.2}, we present a configuration where there are two 3D nulls adjacent to each other, and we observe the QSEB and the chromospheric inverted-Y-shaped jet from the location where the footpoints of their fan surfaces intersect.
We find a high squashing factor at the location of the QSEB for its entire duration. 
A similar scenario of interaction between two adjacent fan surfaces has been studied by \citet{2021SoPh..296...26K} using 3D MHD simulations with the EULAG-MHD code \citep{smolarkiewicz&charbonneau2013jcoph}. 
The initial magnetic field in their simulation is shown to have QSLs in regions where the footpoints of the two fan surfaces interact. During the MHD evolution, self-consistent flows generated by the initial Lorentz forces produce rotational flows around the fan surfaces. This leads to the formation of current sheets due to the misalignment of field lines and triggers torsional fan reconnection \citep{2009PhPl...16l2101P} at the QSL locations. 

The chromospheric inverted-Y-shaped jets in both scenarios of our study resemble in morphology the chromospheric anemone jets described by \citet{2007Sci...318.1591S} observed in the \CaH\ line in SOT/Hinode data. 
These jets were believed to occur as a result of magnetic reconnection between an emerging magnetic dipole and the pre-existing magnetic field. 
The reconnection process in Sect.~\ref{sec: 4.1} appears to follow a similar mechanism, with the chromospheric jet originating at the QSLs formed between the emerging fan and the pre-existing fan-spine topology. The jet fades once the emerging inner fan structure dissipates.
The size of the cusp formed from the merged strands in our examples varies between 1\arcsec -- 2\arcsec, which is similar to the size 1\arcsec -- 3\arcsec\ reported for Hinode chromospheric anemone jets by \citet{2011ApJ...731...43N}. The width of the jets in our examples is approximately 0\farcs3, and the height varies between 1\arcsec --2\arcsec. It has also been speculated that the footpoints of these anemone jets could correspond to EBs \citep{2010PASJ...62..901M}.

%
%
Y-shaped jets on scales comparable to those of our examples have also been reported by \citet{2023Sci...381..867C} in coronal hole EUV observations from Solar Orbiter, and by \citet{2011ApJ...736L..35Y} in the intergranular lanes in the wings of the \Halpha\ line.
They have also been observed in \HeI\ 10830 \AA\ by \citet{2021ApJ...913...59W} who noted bright kernels at the base of the jet which is larger in scale compared to our example. They studied the magnetic topology using nonlinear force-free field (NLFFF) extrapolations and suggested that the magnetic reconnection around QSL associated with a bald patch is the cause of the jet. They also showed that the dome and spire of the jet lie along a region of high squashing factor, consistent with our first example in Sect.~\ref{sec: 4.1}. We also observe similar brightenings, one below the intersection of the jet strands and another at the footpoint of the inner spine of the smaller 3D null.

The jets in this work originate next to the QSEBs and are visible in the blue wing of the \Hbeta\ wavelength, so they could be similar upflows as the RBEs, which are the on-disk counterparts of the Type II spicules.
Figure~\ref{fig:1} shows that the strand associated with the jet discussed in Sect.\ref{sec: 4.1} looks like the spicules in the FOV. From the evolution of this strand, we see that it bends and joins to the spire of the jet.
\citet{2013ApJ...767...17Y} have shown using potential field extrapolations, that some of the RBEs could arise due to magnetic reconnection. They suggest that the RBEs bend above the reconnection site, and also display a brightening below the bending point.
Recently, 
\citet{2025arXiv250405396S} 
demonstrated that a subset of the Type II spicules in their observations are rooted at QSEB locations. As a result, QSEBs and spicules reflect the conversion of magnetic energy into thermal and kinetic energy, respectively.

To conclude, in this paper, we have demonstrated, through observational and modelling evidence, how small-scale dynamic phenomena--such as QSEBs, UV brightenings, and chromospheric inverted-Y-shaped jets--are interconnected and arise from a common magnetic reconnection scenario of interacting fan-spine topologies.

\begin{acknowledgements}
We thank Guillaume Aulanier, Reetika Joshi and Carlos José Díaz Baso for the helpful discussions. 
A.P. also thanks Sanjay Kumar for the helpful discussions.
The Swedish 1-m Solar Telescope (SST) is operated on the island of La Palma by the Institute for Solar Physics of Stockholm University in the Spanish Observatorio del Roque de los Muchachos of the Instituto de Astrof{\'\i}sica de Canarias.
The SST is co-funded by the Swedish Research Council as a national research infrastructure (registration number 4.3-2021-00169).
This research is supported by the Research Council of Norway, project number 325491, 
and through its Centres of Excellence scheme, project number 262622. 
IRIS is a NASA small explorer mission developed and operated by LMSAL, with mission operations executed at NASA Ames Research Center and major contributions to downlink communications funded by ESA and the Norwegian Space Agency.
SDO observations are courtesy of NASA/SDO and the AIA science teams.
%
J.J. is grateful for travel support under the International Rosseland Visitor Programme. 
J.J. acknowledges funding support from the SERB-CRG grant (CRG/2023/007464) provided by the Anusandhan National Research Foundation, India.
A.P. and D.N.S acknowledge support from the European Research Council through the
Synergy Grant number 810218 (``The Whole Sun'', ERC-2018-SyG).
D.N.S. also acknowledges the computer resources at the MareNostrum
supercomputing installation and the technical support provided by the
Barcelona Supercomputing Center (BSC, RES-AECT-2021-1-0023,
RES-AECT-2022-2-0002), as well as the resources provided by 
Sigma2 - the National Infrastructure for High Performance Computing and Data Storage in Norway.
We acknowledge using the visualisation software VAPOR (www.vapor.ucar.edu) for generating relevant graphics.
We made much use of NASA's Astrophysics Data System Bibliographic Services.
\end{acknowledgements}

\bibliographystyle{aa}
\bibliography{aditis_ref} 

\end{document}